\documentclass[journal,twoside]{IEEEtran}
\usepackage{xcolor}
\usepackage{graphicx}
\usepackage[colorlinks, citecolor=blue]{hyperref}
\usepackage{array}
\usepackage{cite}
\usepackage{subfig}
\usepackage{amsmath,amssymb,amsfonts}
\usepackage{graphicx}
\usepackage{textcomp}
\usepackage{multirow}
\usepackage{mathtools}
\usepackage{soul}
\usepackage{algorithm}

\usepackage{algpseudocode}

\usepackage[T1]{fontenc}%
\usepackage[utf8]{inputenc}%
\usepackage{lmodern}%
\usepackage{textcomp}%
\usepackage{lastpage}%
\usepackage{graphicx}%
\usepackage{multicol}

\newcolumntype{C}[1]{>{\centering\let\newline\\\arraybackslash\hspace{0pt}}m{#1}}
\newcolumntype{L}[1]{>{\raggedright\let\newline\\\arraybackslash\hspace{0pt}}m{#1}}

\begin{document}
\include{review}
\newpage

\title{\huge{A Semi Black-Box Adversarial Bit-Flip Attack with Limited DNN Model Information}}
\author{
Behnam Ghavami\textsuperscript{1}, Mani Sadati, Mohammad Shahidzadeh\textsuperscript{2}, Lesley Shannon\textsuperscript{2}, Steve Wilton\textsuperscript{1}

\textsuperscript{1}University of British Columbia, Canada\\ \textsuperscript{2}Simon~Fraser~University, Canada\\
\thanks{
} 
This paper has been accepted for presentation at IEEE International Conference on Computer Design, 2024 (ICCD 2024)
}

\markboth{}%
{}


\pagenumbering{gobble}
\maketitle

\vspace{-0.5in}

\begin{abstract}
Despite the rising prevalence of deep neural networks (DNNs) in cyber-physical systems, their vulnerability to adversarial bit-flip attacks (BFAs) is a noteworthy concern. This paper proposes B3FA, a semi-black-box BFA-based parameter attack on DNNs, assuming the adversary has limited knowledge about the model. We consider practical scenarios often feature a more restricted threat model for real-world systems, contrasting with the typical BFA models that presuppose the adversary's full access to a network's inputs and parameters. The introduced bit-flip approach utilizes a magnitude-based ranking method and a statistical re-construction technique to identify the vulnerable bits.
We demonstrate the effectiveness of B3FA on several DNN models in a semi-black-box setting. For example, B3FA could drop the accuracy of a MobileNetV2 from 69.84\% to 9\% with only 20 bit-flips in a real-world setting.
\end{abstract}

\section{Introduction}\label{sec:intro}

In recent years, deep neural networks (DNNs) have achieved tremendous success in different computer vision tasks such as image classification, object detection, and segmentation.
As DNNs become more popular and applicable in real-world scenarios, their security and safety issues are also becoming crucial; e.g., in self-driving cars or medical devices, even a single malfunction can have dire consequences. In particular, security attacks exploiting hardware-level flaws in memory elements can be challenging for edge devices that process sensitive data \cite{kim2014flipping}.

Fault injection attacks (FIAs) belong to a class of hardware-oriented attacks where a fault is deliberately injected into hardware for malicious purposes. Recently, several FIAs targeting DNNs have emerged, including adversarial parameter bit-flip attacks (BFAs), which aim to disrupt a model's functionality by perturbing its vulnerable parameters \cite{yao2020deephammer, rakin2019bit}. This attack process typically involves two steps. In the first step, the attacker employs side-channel attacks to partially extract model information. Then, in the second step, she may continue by applying gradient-based bit-flipping techniques to the partially recovered model. Previous adversarial parameter attacks assumed white-box settings, wherein the adversary has full access to the network architecture, all weight values (full knowledge of the weight bits), and a portion of the inference dataset. However, real-world threat models are often more restricted, and the assumed information is not available in most situations \cite{rakin2021deepsteal}.



\begin{figure}[!tb]
	\begin{center}
	    \vspace{-0.05in}
		\includegraphics[width = 0.45\textwidth]{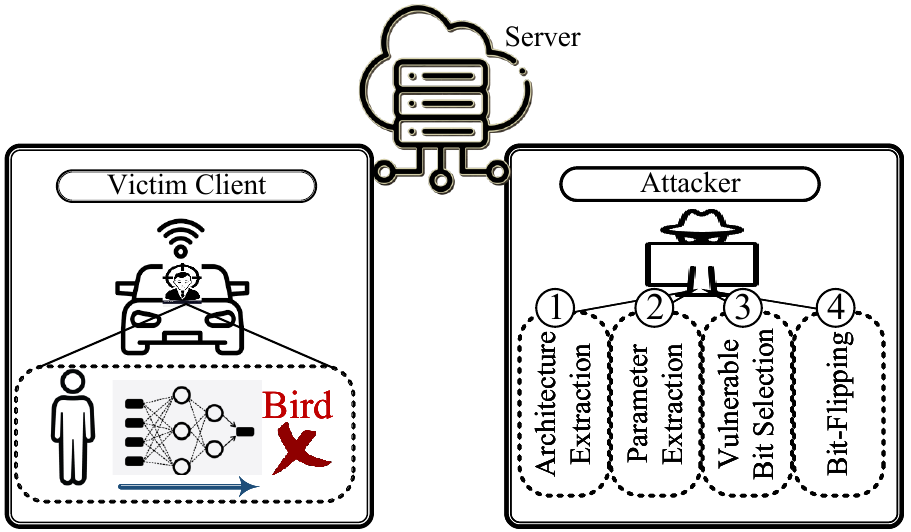}
		\vspace{-0.1in}
		\caption{ An overview of B3FA, in which the attacker and the victim client use a same service environment. 
}
		\vspace{-0.27in}
		\label{fig:flow}
	\end{center}	
\vspace{-0.01in}
\end{figure}

In this paper, we introduce a practical semi-black box adversarial bit-flip attack (B3FA), which is able to destroy the DNN functionality via resembling a real-world scenario where the adversary has limited knowledge of the model to be attacked. 
In the proposed flow, shown in Fig. \ref{fig:flow}, the attacker could extract the model's architecture via the traditionally side-channel attack technique. Then, she takes advantage of the architectural information to partially recover the retrievable parameters. The attacker finds the vulnerable bits from the \textbf{\textit{partially recovered model}} using a novel technique without any access to training data. Finally, the attacker flips the vulnerable bits in the memory to crush the victim's DNN functionality. 

To evaluate the efficacy of B3FA, we conducted extensive practical experiments using popular network structures such as MobileNetV2, VGG16, and ResNet50, as well as different weight representation formats including 8bit, 6bit, and 4bit quantized parameters. The experiments were carried out on CIFAR-10 and CIFAR-100 datasets to assess the performance of B3FA in diverse scenarios.
Experiments are done with different DRAM memories and parameter recovery rates to investigate various parameter extraction settings. The results demonstrate that an attacker is able to malfunction all the DNN models with only a very small amount of bit-flips (roughly 0.00001\% of parameter bits). For example, an 80\% extracted MobileNetV2 network was destroyed using only 20 bit-flips, out of 28 million bits, causing an accuracy degradation from 69.84\% to 9\%. The accuracy of the same model with 60\% extracted parameters will be degraded down to 50\% after 20 bit-flips.

To the best of our knowledge, we are the first to propose a semi-black-box attack flow for BFAs on DNNs. The main contributions of this paper are as follows:
\begin{itemize}
    \item We demonstrate how existing DNN architecture recovery techniques yield partially recovered models, emphasizing the impracticality of traditional BFAs, which assume fully perfect model extraction.

    \item We propose a magnitude-based vulnerable bit ranking approach, called normalized filter-wise $\mathcal{L}_2$-norm ranking (FL2R), to find the most vulnerable and sensitive bits in the fully recovered model without having access to training/inference dataset.
    \item We extend FL2R to partially recovered DNNs via introducing a novel statistical method called closer to zero reconstruction (CZR) to predict unextracted weights. 
    \item We have experimentally validated the proposed B3FA flow on DNN-based image classification in a real-world setting.
    
\end{itemize}


The rest of this paper is as follows: Section \ref{sec:related} gives an overview of prior DNN attacks, and Section \ref{sec:threat} explains the threat model. Section \ref{sec:proposed} presents the proposed attack. Sections \ref{sec:setup} and \ref{sec:results}, discuss the experimental setup and results. Finally, Section \ref{sec:disc&con} gives a discussion about the potential defense methods and the feasibility of performing B3FA having a partially extracted DNN architecture and concludes the paper.
\section{Related Work}
\label{sec:related}







The most well-known fault injection attacks on dynamic random access memory (DRAM) are rowhammer attacks \cite{kim2014flipping}, \cite{agoyan2010flip}. They provide the attacker with a profile of the vulnerable bits stored in the main memory (i.e. DRAM) and can flip any bit of any given target address. Early attempts at exploiting rowhammer to attack DNN parameters focus on flipping the most significant bits (MSB) of network weights or biases \cite{liu2017fault}.However, these attacks are limited to a model with high precision (i.e., 32-bit floating point) parameters and fail in quantized DNNs that are more noise-resistant \cite{rakin2019bit,hong2019terminal}.

A key step forward in the adversarial weight attack and BFA is introduced in \cite{rakin2019bit}, where 8-bit fixed-point quantized networks are under attack. Using all weight bit information, this attack gradually decreases the DNN accuracy by finding and flipping the most vulnerable weight bits iteratively. BFA was practically explored by attacking victim information through rowhammer vulnerability characteristic \cite{yao2020deephammer}. 
Lately, Zhao et al. \cite{zhao2019fault} introduced a bit flipping attack on a DNN classifier in order to stealthily misclassify a few predefined inputs. 
Also, Ghavami et al. \cite{ghavami2022stealthy} presented a stealty attack on DNNs to misclassify crafted inputs but reserve the classification accuracy for clean inputs via smart bit flipping. Recent works can perform a targeted bit flip attack on full precision models \cite{zhao2019fault}, although they require large amounts of weight perturbation.
Rakin et al. \cite{rakin2021t} introduced an adversarial bit flip attack on quantized DNN models where the main goal is to identify the weights that are highly associated with the misclassification of a targeted output. Lately,  Bai et al. \cite{bai2021targeted} developed a binary integer programming formulation for targeted adversarial bit flipping.
Recently, a data-independent version of BFA using distilled data was introduced \cite{ghavami2021bdfa}. But, this attack still needs the full information of architecture and weight bits to generate distilled data.




\textbf{Limitations of previous works:} 
To facilitate a comprehensive comparison between the previous bit-flip attacks and the proposed B3FA, we present a detailed analysis of their assumptions in Table \ref{tab:assumptions}. All prior bit-flip attacks require access to full information of weight bits and a portion of training/inference dataset (i.e white box threat model), where such requirements may not be applicable in all scenarios.

\begin{table}[tb!]
\begin{center}
\vspace{0.1in}
\vspace{-0.05in}    
\scalebox{0.72}{
\begin{tabular}
{|l|c|c|c|c|}
    \hline 

     && \multicolumn{3}{|c|}{Assumptions} \\   \cline{3-5}
     & \shortstack{Bit-Ranking} &\shortstack{Input data}&\shortstack{Parameter\\Extraction}&\shortstack{Architecture\\Extraction}  \\   \hline
    \cline{1-5} BFA\cite{rakin2019bit} & Gradient & \centering One batch & Full & Full  \\
    \cline{1-5} T-BFA\cite{rakin2021t} & Gradient & \centering One batch & Full & Full  \\
    \cline{1-5} DeepHammer\cite{yao2020deephammer} & Gradient & \centering One batch & Full & Full  \\
    \cline{1-5}  SBFA\cite{ghavami2022stealthy} & Gradient & \centering One batch & Full & Full \\ 
    \cline{1-5}  TA-LBF\cite{bai2021targeted} & Gradient & \centering One batch & Full & Full \\  
    \cline{1-5}  BDFA\cite{ghavami2021bdfa} & Gradient & \centering None & Full & Full \\  
    \cline{1-5} \textbf{B3FA (ours)} & Scaled L2-norm & \centering None & Partially & Partially  \\  
    
    \hline
    
\end{tabular}
}
\vspace{-0.05in}
\caption{Comparison of adversarial parameter attack methods and assumptions of their threat models.}

    \label{tab:assumptions}
\vspace{-0.3in}
\end{center}
\end{table}

\section{THREAT MODEL}
\label{sec:threat}




We consider an attacker who aims to crush the functionality of a victim DNN application. We assume the attacker has access to a user-space un-privileged process that runs on the same machine as the victim DNN service, a resource-sharing environment offering machine learning inference service (which is a popular application paradigm due to the prevalence of MLaaS platforms \cite{ribeiro2015mlaas}). In this environment, a read-only memory is shared between the attacker and the victim DNN through library sharing or memory deduplication features.
Due to the co-location of victim and attacker processes, when a shared cache is used between multiple cores, it can lead to computation leakage at the cache level.
The attacker mainly exploits remote-based side-channel attacks, e.g. cache attack, to extract the DNN architecture either completely or partially \cite{hong20200wn,hu2020deepsniffer}. Moreover, we assume she is able to run fault injection-based side-channel attack to partially extract the parameters via deploying the well-known software-based rowhammer bit-flip injection \cite{rakin2021deepsteal,kwong2020rambleed,razavi2016flip,seaborn2015exploiting}. 
Lastly, it is assumed that the attacker is aware of the representation format in which the parameters are stored in memory, such as the commonly used two's-complement format for storing integer (quantized) values. If the representation format is unknown, the attackers can perform multiple sets of attacks to ensure they cover the few commonly used representations. They can prioritize the more popular representations which are used for both high accuracy and high robustness such as quantized 8-bit integers.

B3FA is the first adversarial bit-flip attack with the \textbf{\textit{semi-black-box setting}},  where there is no prior knowledge of DNN architecture and parameters. In addition, there is no access to any portion of the training or inference dataset.

\section{Semi-Black-Box Adversarial Bit-Flip Attack}\label{sec:proposed}



\begin{figure*}[!tb]
	\begin{center}
	    \vspace{-0.4in}
		\includegraphics[width = 1\textwidth]{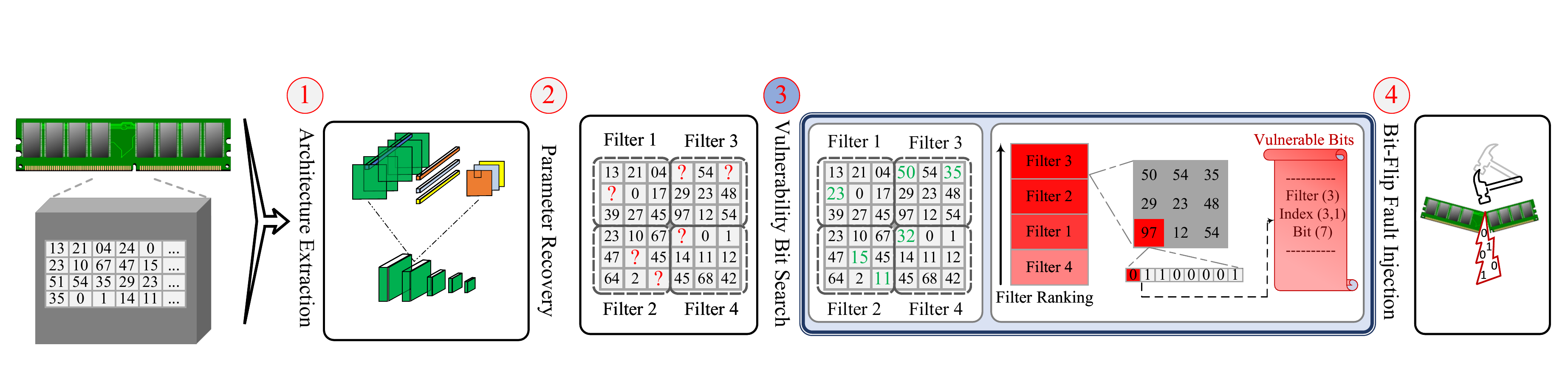}
		\vspace{-0.3in}
		\caption{
		Four main stages of B3FA framework.
		The attacker begins by extracting the architecture \raisebox{.5pt}{\textcircled{\raisebox{-.9pt} {1}}} and partial parameters \raisebox{.5pt}{\textcircled{\raisebox{-.9pt} {2}}} from a DRAM that contains the victim's information. Then, she discovers a list of candidate bits using the \textit{proposed blind vulnerable bit search} \textcolor{blue}{\raisebox{.5pt}{\textcircled{\raisebox{-.9pt} {3}}}}, which she will use to flip in the final step \raisebox{.5pt}{\textcircled{\raisebox{-.9pt} {4}}}. }
		\vspace{-0.27in}
		\label{fig:quandequan}
	\end{center}	
\vspace{0.05in}
\end{figure*}

Fig. \ref{fig:quandequan} shows the main stages of the proposed B3FA flow over a deep learning model, $M(A,\theta)$, with the architecture information of $A$ and parameters of $\theta$. The attacker utilises an architecture extraction in \textit{stage 1} to uncover the DNN structure ($A'$) from the memory data. 
In \textit{stage 2}, she leverages the structure information to partially recover some of the DNN parameters ($\theta'$) using a crude but practical extraction strategy. In \textit{stage 3}, the attacker uses our novel magnitude-based importance evaluation technique and statistical reconstruction method to discover the most vulnerable bits in the partially recovered model. In the last stage, the attacker uses a double-sided rowhammer attack to tamper with the memory cells on the victim page, resulting in an incorrect answer from the ML environment to the victim client in filed operation. Each stage will be explained in detail in the following sub-sections.

\subsection{Architecture Extraction}

In this stage, the attacker aims to extract the exact DNN architecture from the memory (i.e. $A' = A$). The information includes specifications of each layer such as weight filter dimensions, which would be used to determine the in-memory layout of parameters in stage 2.

Side-channel attacks may be used to leak information about the model architecture \cite{yan2020cache}, \cite{hu2020deepsniffer}, \cite{batina2019csi}, \cite{hong20200wn}. Model structure could be extracted through micro-architectural attacks \cite{yan2020cache}, physical side-channel attacks \cite{batina2019csi}, and bus snooping attacks \cite{hu2020deepsniffer}. Most of the existing side-channel attacks can obtain high-level model information (e.g. model architectures) with a very high accuracy. Without loss of generality, in this paper, we use cache side-channel attacks (Flush+Reload) \cite{hong20200wn} in order to extract the DNN architecture in our threat model. 



\subsection{Partial Parameter Recovery}



Having the extracted model architecture $A'$, the adversary performs an extraction algorithm which outputs the recovered parameters $\theta'$. Parameter extraction is an emerging field that tries to steal detailed information of a deep learning model from a victim device. The attacker deploys a memory fault injection technique to recover DNN parameters \cite{rakin2021deepsteal}, \cite{breier2021sniff}. The double-sided rowhammer fault injection technique could be applied as the information leakage vector. 

Although parameter extraction methods can reveal the details of the model parameters, the extraction is only \textit{partial}; i.e. the parameters are not fully recovered and some of the weight bits are left unextracted ($\theta' \neq \theta$), particularly in practical settings. This is due to the fact that the exact model recovery, where the adversary knows the exact value of each parameter in the model, is practically impossible \cite{jagielski2020high}. We define model \textit{recovery rate}, expressed as a percentage, as the fraction of parameters exposed via an extraction procedure in an environment setting.

\subsection{Gradient-based BFA Deployment Challenge}\label{sec:challenge}
\begin{figure}[!tb]
	\begin{center}
	    \vspace{-0.01in}
		\includegraphics[width = 0.5\textwidth]{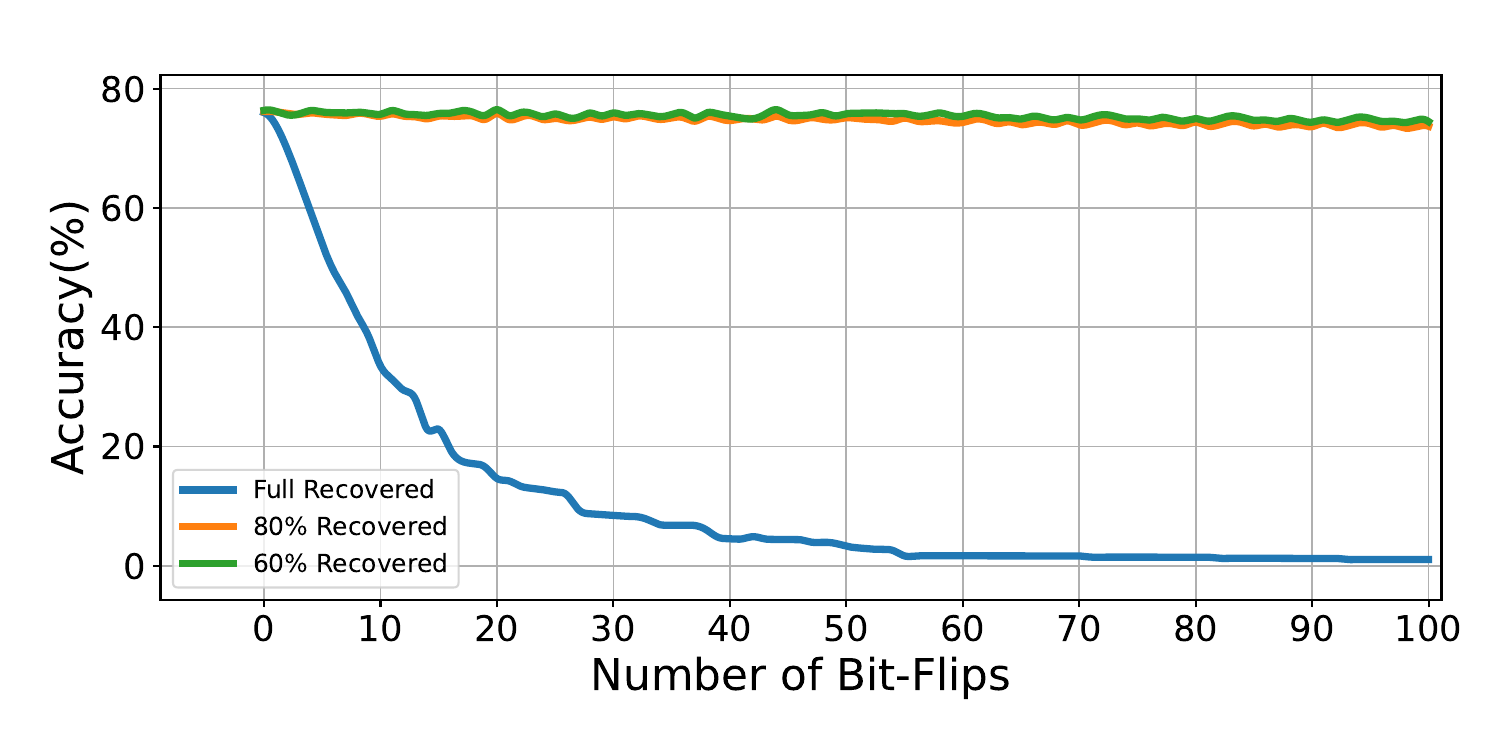}
		\vspace{-0.27in}
		\caption{Examination of traditional BFA deployment on partially restored DNNs.}
		\vspace{-0.15in}
		\label{fig:bfa-failed}
	\end{center}	
\vspace{-0.0in}
\end{figure}

Since one can only partially extract the model parameters, conventional bit-flip attacks---which are based on gradient descent computation---cannot be successfully deployed to crush the functionality of these partially recovered models $M(A',\theta')$.
To better demonstrate this issue, we ran BFA at various recovery rates on a DNN (details in Section \ref{sec:white-box}), and the results are shown in Fig. \ref{fig:bfa-failed}. As can be seen, BFA is only successful in the fully recovered model, but fails to reduce the model accuracy in the partially recovered setting. 

This problem arises from two main issues. \textit{Issue 1:} Conventional BFAs need full parameter information to compute loss gradients and locate vulnerable weight bits. \textit{Issue 2:} Gradient computation requires a portion of the training/inference dataset, which is inaccessible in a real threat model.

\subsection{Proposed Blind Vulnerable Bit Search}
In this section, a magnitude-based vulnerable bit search is first proposed to find the most vulnerable bits of parameters in a fully recovered DNN model, i.e. $M(A , \theta)$. Unlike the gradient-based vulnerable bit search, this blind search method does not require any portion of the training/inference dataset (addressing Issue 2 in Section \ref{sec:challenge}). 
Then, we extend the proposed method for partially recovered parameter model, $M(A,\theta')$, using a statistical reconstruction technique (addressing Issue 1 in Section \ref{sec:challenge}).



\subsubsection{Normalized Filter-wise L2-norm Ranking (FL2R) for Fully-Recovered DNNs}
Here, we propose a magnitude-based approach, shown in Algorithm \ref{alg:euclid1}, to estimate the vulnerability of each filter in the convolutional layers of DNNs. Inspired from some of the importance estimation methods in pruning DNNs \cite{he2018soft}, \textit{we score the vulnerability of each filter based on the normalized $\mathcal{L}_2$-norm (Frobenius norm) of the weights in that filter.} Then, we use these scores to find the most vulnerable filters and perturb their most sensitive bits.

Generally, in DNNs, filters with higher $\mathcal{L}_2$-norm values ($\mathcal{L}_p$-norm in general) lead to higher activation values in their corresponding output feature maps \cite{he2018soft}, and higher activation values lead to a higher numerical impact on the final output of DNN models. Hence, filters with higher $\mathcal{L}_2$-norm are more important in the accurate functionality of the model. The $\mathcal{L}_2$-norm of each filter is calculated as below:
\begin{equation}
    ||\mathcal{F}_{i}^{(l)}||_{2} = \sqrt{\sum_{c=1}^{C_{in}^{(l)}}\sum^{K^{(l)}}_{k_1=1}\sum^{K^{(l)}}_{k_2=1} |W_{i,c,k_1,k_2}^{(l)}|^{2}}
\end{equation}
where the ${l}^{th}$ convolutional layer contains $C_{out}^{(l)}$ weight filters $F_{i}^{(l)}$ $(0 \leq i < C_{out}^{(l)})$. Each filter has a shape of $C_{in}^{(l)} * K^{(l)} * K^{(l)}$, where $C_{out}^{(l)}$ and $C_{in}^{(l)}$ are the output and input channel size, and $K^{(l)}$ is the kernels' height and width in the $lth$ layer. The parameters of layer $l$ can also be displayed as a weight matrix of $W^{(l)} \in \mathcal{R}^{C_{out}^{(l)} * C_{in}^{(l)} * K^{(l)} * K^{(l)}}$.

\begin{figure}[!tb]
	\begin{center}
	    \vspace{-0.1in}
		\includegraphics[width = 0.5\textwidth]{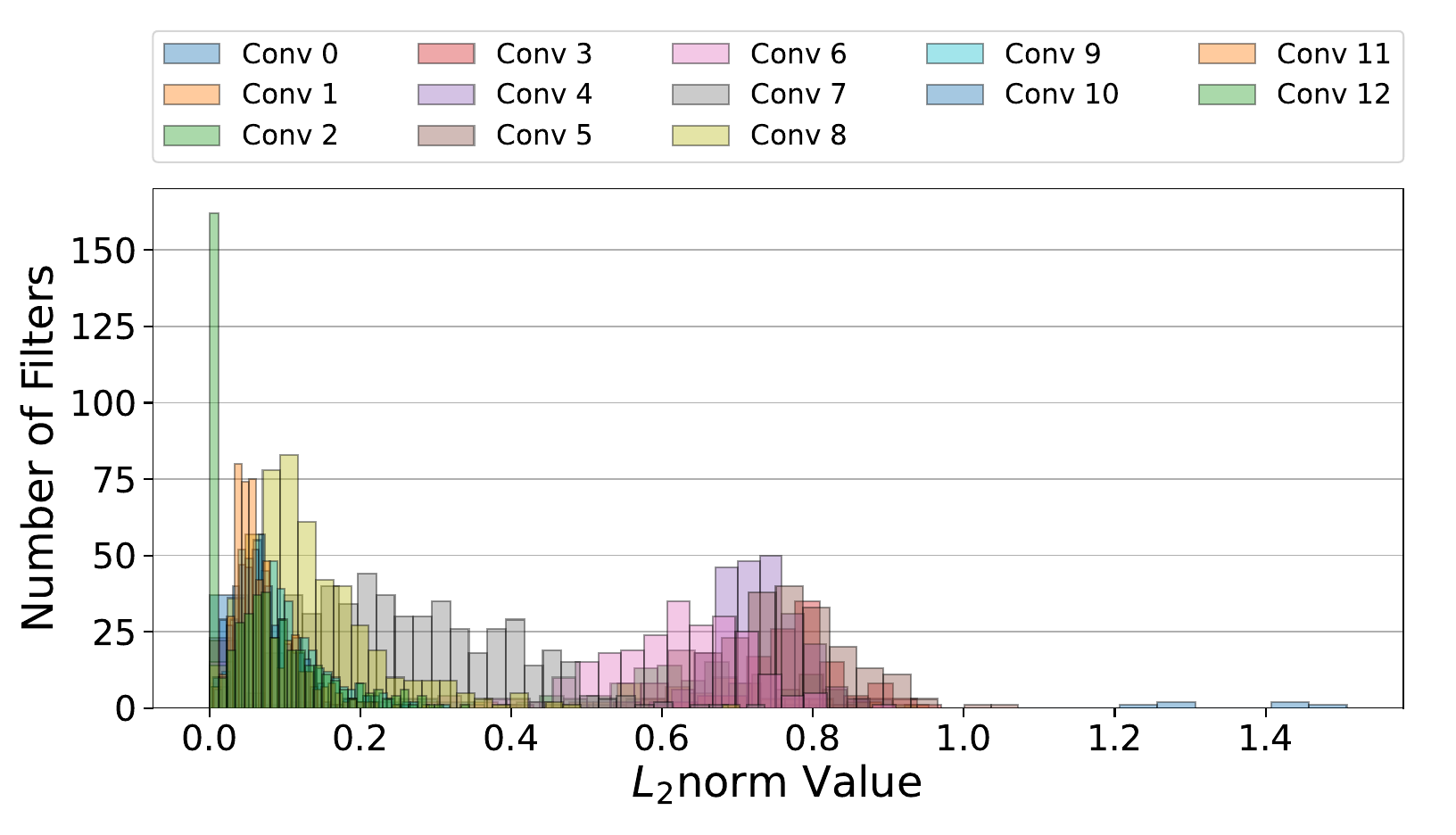}
		\vspace{-0.25in}
		\caption{$\mathcal{L}_2$-norm of different filters across different layers of the
VGG-16 trained on CIFAR-10.}
		\vspace{-0.1in}
		\label{fig:distribfilter}
	\end{center}	
\vspace{-0.2in}
\end{figure}


Fig. \ref{fig:distribfilter} illustrates the distribution of $\mathcal{L}_2$-norm values of the filters for each convolutional layer in a VGG-16 network trained on the CIFAR-10 dataset. It shows that, filters in different layers have different $\mathcal{L}_2$-norm distributions; for example, most of the filters in layer $conv6$ (with $C_{in}=256$) have an $\mathcal{L}_2$-norm of around $0.6$, while most filters in $conv2$ (with $C_{in}=64$) have an $\mathcal{L}_2$-norm close to $0.1$. This is due to the fact that $\mathcal{L}_2$-norm of each filter directly relates to the size of that filter, and each layer has a different filter size. Therefore, in order to compare different layers fairly, we propose normalizing the importance score of filters with respect to their sizes. The vulnerability of each filter ($\mathcal{I}(F_{i}^{(l)})$) is calculated as the normalized $\mathcal{L}_2$-norm calculated as below:
\begin{equation}
    \mathcal{I}(F_{i}^{(l)}) = \frac{\sqrt{\sum^{C_{in}^{(l)}}_{c=1}\sum^{K^{(l)}}_{k_1=1}\sum^{K^{(l)}}_{k_2=1} |W_{i,c,k_1,k_2}^{(l)}|^{2}}}{C_{in}^{(l)}*K^{(l)}*K^{(l)}} 
\end{equation}
Based on the proposed metric, we locate the most vulnerable filter, $F^*$, in the first step of the Algorithm \ref{alg:euclid1} (line 3).




In order to disrupt and neutralize the chosen filter, we need to decrease $\mathcal{L}_2$-norm as much as possible by selecting the parameter with the highest absolute value, $w^*$ (line 4 in Algorithm \ref{alg:euclid1}). Here, we assume the parameter with the highest absolute value is in the range of $[2^{N_q-1} - 2^{N_q-3}, 2^{N_q-1}]$ (the top 25\% quantile of a $N_q$-bit quantization). This is a valid assumption since the quantization range is determined based on the parameters with the highest value \cite{gholami2021survey}. Therefore, the most effective bit-flip position is the sign bit (($N_q$-1)-\textit{th} bit) since it causes a $2^{N_q-1}$ unit shift and changes the value in the range of $[2^{N_q-1} - 2^{N_q-3}, 2^{N_q-1}]$ to $[0 , 2^{N_q-3}]$, which is the closest interval to zero by flipping only one bit (lines 5 and 6 in Algorithm \ref{alg:euclid1}). Finally, we recalculate the vulnerability score of $F^*$. The whole process of finding vulnerable bits is then repeated enough times to compute the top $N_{bf}$ vulnerable bits. $N_{bf}$ is an adjustable parameter that the attacker configures as the desired number for bit-flip injections based on the need.


\begin{algorithm} [!t]
\caption{Vulnerable Bit Selection}\label{alg:euclid1}
\textbf{Input:} DNN parameters after reconstruction $W$; Number  \hspace{10mm}of Vulnerable Parameters Required $N$;\\
\textbf{Output:} $VB$ = Top $N_{bf}$ vulnerable weight bits\\
\textbf{Initialize:} $VB = \{\}$;

\begin{algorithmic}[1]
\Procedure{Magnitude-Based Bit Selection}{}
    \While{Size($VB$) $<$ $N_{bf}$}
        \State $F^*$ $\gets$ $argmax_F$ $\mathcal{I}(F_{n}^{(l)})$
        \State $w^*$ $\gets$ $argmax_w$ $w_{c,k_1,k_2}^2; w \in F^*$ 
        \State $VB$.Add($w^*$)
        \State $w^*$ $\gets$ $Flip_{MSB}(w^*)$
        \State Recalculate $\mathcal{I}(F^*)$
    \EndWhile \label{euclidendwhile}
    \State \textbf{return} $VB$
\EndProcedure
\end{algorithmic}
\label{Alg:main}
\end{algorithm}
\setlength{\textfloatsep}{6pt}

\subsubsection{FL2R on Partially-Recovered DNNs using Statistical Reconstruction}\label{sec:CZR}
\begin{figure}[!tb]
	\begin{center}
	    \vspace{-0.23in}
		\includegraphics[width = 0.4\textwidth]{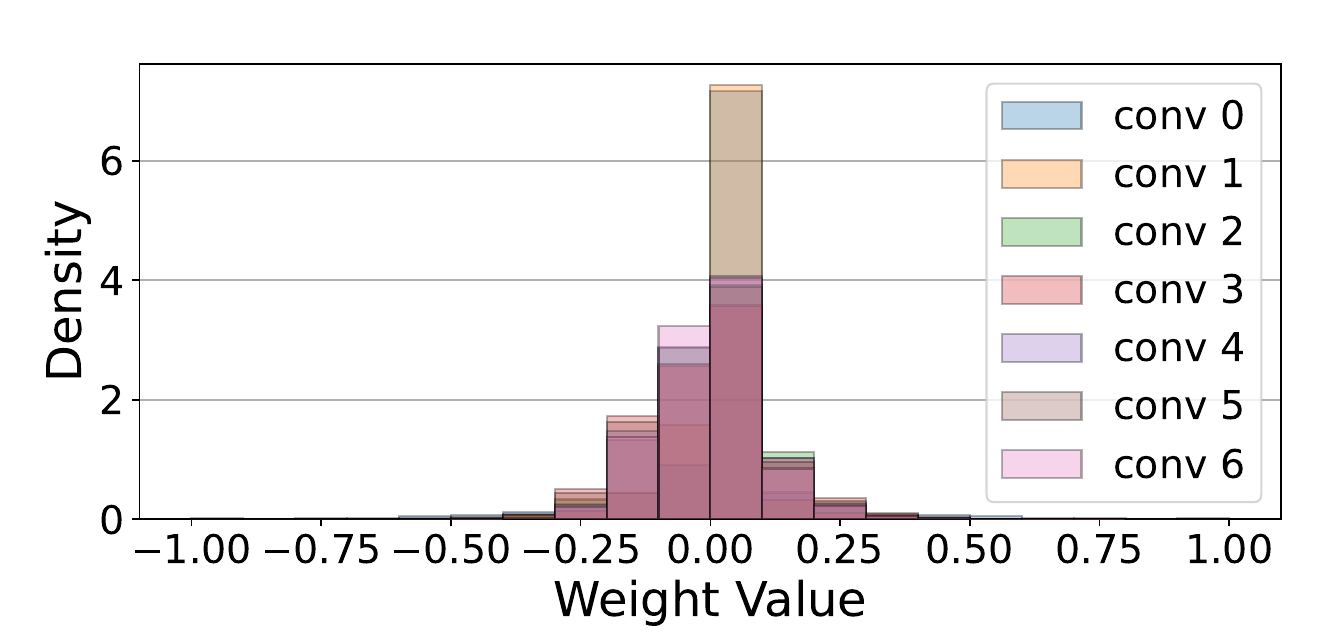}
		\vspace{-0.1in}
		\caption{Distribution of weights across different layers.}
		\vspace{-0.1in}
		\label{fig:distributionvgg}
	\end{center}	
\end{figure}

In this section, we extend FL2R on partially-recovered models. Since FL2R requires full information of parameters to compute the $\mathcal{L}_2$-norm of each filter, we propose a statistical reconstruction method to determine the unrecovered weight bits. \textit{Based on statistical information of DNN parameters, we predict each unrecovered parameter bit in a way that the parameter's value gets as closer as possible to zero.} Generally, the weight distribution of convolutional layers is bell-shaped such as Gaussian or Laplacian \cite{lin2016fixed}. Fig. \ref{fig:distributionvgg} shows these distributions in a VGG-16 network trained on the CIFAR-10 dataset. It indicates that the probability of the value of a weight being closer to zero is much higher than having a high absolute value. That is, most of the weights are clustered around zero while few of them are spread in a long tail. Hence, we introduce closer to zero reconstruction (CZR) to reconstruct each DNN parameter as closer to zero. CZR tries to predict the unrecovered bits of each weight such that the absolute value of the weight be as low as possible. 


In our threat model, we assumed the quantized parameters are stored in memory using the two's-complement format (which is the most popular format for storing integer values in memory). This means that the smallest positive number is 0 or 00000000 (for $N_q = 8$) and the biggest negative number is -1 or 11111111 (for $N_q = 8$). Therefore, to reconstruct the parameters, we categorize them into three categories: 1) the parameters with the recovered sign bit of 0 (a positive weight), 2) the parameters with the recovered sign bit of 1 (a negative weight), and 3) the parameters with an unrecovered sign bit. For a positive parameter, all of the un-extracted bits in that parameter will be set to zero. For a negative parameter, all unknown bits will be set to one. For parameters with the unrecovered sign bit, we consider the sign bit once positive and once negative and reconstruct the other unknown bits with respect to the assigned sign bit. Then, by comparing the two computed values, we consider the one with lower absolute value as the reconstructed parameter.

\subsection{Precise Memory Fault Injection}
Having the vulnerable bit list provided in stage 3, the adversary aims to precisely inject bit-flips into the computing memory for DNN inference. Precise rowhammering has been shown to be effective for bit-flip injection at specific targeted locations. Especially, double-sided rowhammer-based bit-flip attacks \cite{razavi2016flip}, such that the victim page is placed as the vulnerable row and two attacker pages are placed as the aggressors, could be utilized. We deploy the Deephammer technique \cite{yao2020deephammer} to precisely bit-flip the victim's DNN parameters in the memory.

\section{Experimental Setup} \label{sec:setup}
\subsection{DNN Architectures and Datasets}
The experiments use three DNN models: MobileNetV2 \cite{sandler2018mobilenetv2}, VGG16 \cite{simonyan2014very}, and ResNet50 \cite{he2016deep}, trained on CIFAR-10 and CIFAR-100 \cite{CIFAR-10}. Table \ref{tab:baseline} shows their baseline accuracy. We apply layer-wise symmetric uniform 8-bit quantization to the DNN parameters \cite{gholami2021survey}.

\begin{table}[htb!]
\begin{center}
\scalebox{1.0}{
\begin{tabular}
{|l|r|r|r|}
    \hline 

    &  \multicolumn{1}{c|}{ResNet50}   & \multicolumn{1}{c|}{VGG16}  & \multicolumn{1}{c|}{MobileNetV2} \\ \hline 
    \cline{1-4} CIFAR-10 (\%) & 94.63\% & 93.06\% & 91.6\%   \\ 
    \cline{1-4}  CIFAR-100 (\%) & 75.96\%  & 72.34\% & 69.84\% \\  
    
    \hline
    
\end{tabular}
}
\vspace{-0.01in}
\caption{Baseline Accuracy of ResNet50, VGG16 and MobileNetV2 on CIFAR-10 and CIFAR-100 datasets.}

    \label{tab:baseline}
\vspace{-0.2in}
\end{center}
\end{table}

\begin{figure*}[!ht]
	\begin{center}
		\vspace{-0.4in}
		\includegraphics[scale=0.3]{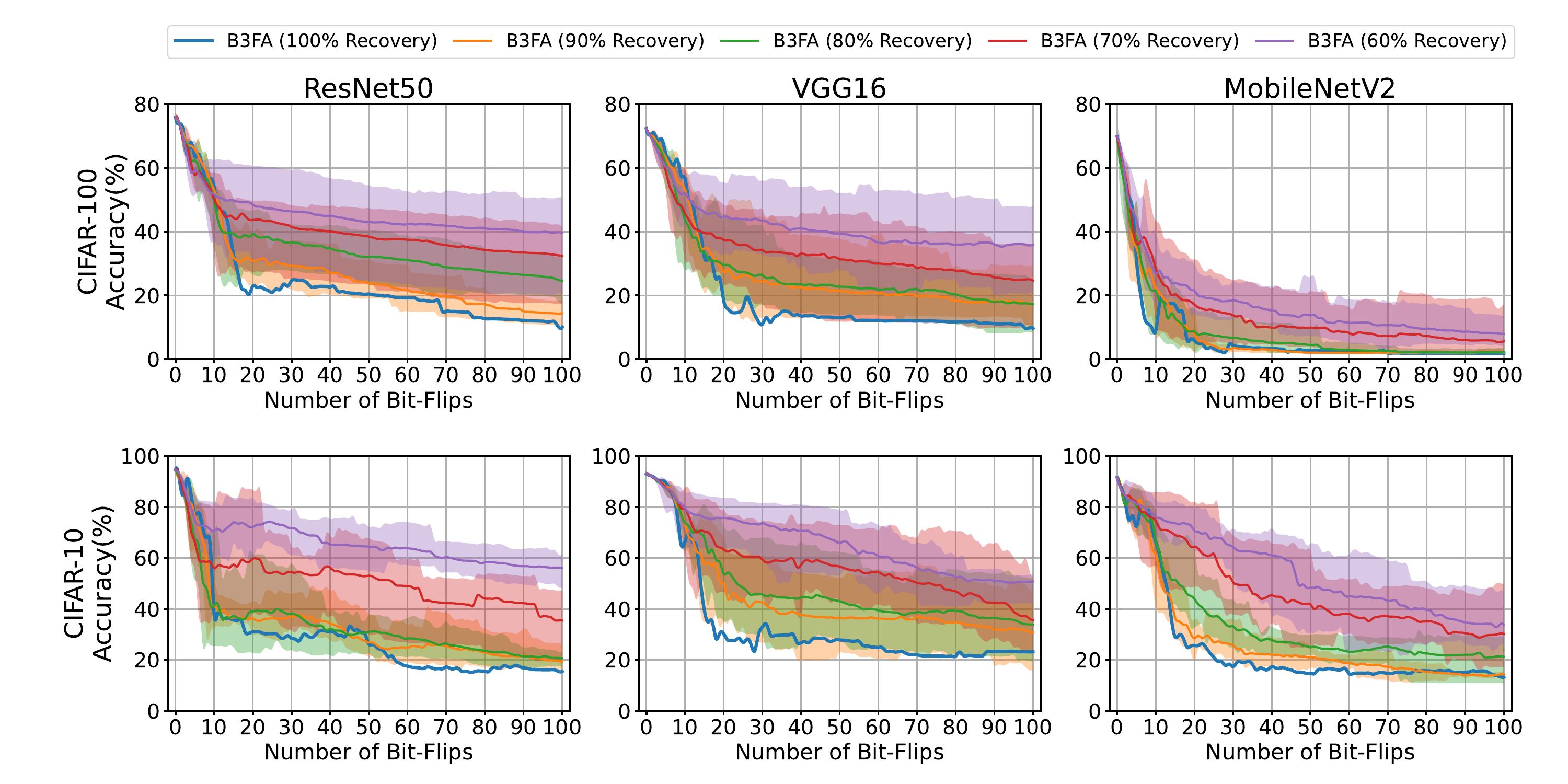}
		\vspace{-0.1in}
		\caption{Model accuracy after each iteration of B3FA with different recovery rates on ResNet50, VGG16 and MobileNetV2, trained on CIFAR-10 and CIFAR-100. We run ten experiments on different DRAMs for each model, and the region in shadow indicates the error-band.}
		\label{fig:aveAcc}
		\vspace{-0.3in}
	\end{center}	
\end{figure*}

\subsection{System Setup} \label{subsec:setup}
Experiments are based on the threat model in Section \ref{sec:threat}, assuming a DL system deployed in a resource-sharing MLaaS environment. We use the PyTorch \cite{paszke2019pytorch} DL framework with the FBGEMM backend, on Ubuntu 20.04 LTS with a 5.13.0-40-generic kernel. The hardware includes an Intel i5-4590 3.30 GHz (Haswell) with 4 cores, 6MB shared cache, and 8GB DDR3 memory. To account for different DRAM leakage positions, we use 10 different DDR3 memory devices for the experiments.

\subsection{Architecture and Parameter Extraction} \label{subsec:setup-arch-param}
In order to extract the DNN architecture we used cache side-channel attacks (Flush+Reload) as described in \cite{hong20200wn}. First, we analyze the DL framework to identify the starting line of each operation. In the second step, the Flush+Reload attack is utilized. The cache line, which contains the data, is flushed corresponding to the target address, such that the data is evicted from the cache and only exists in memory. Then, we wait until the victim has enough time to call the targeted function. Finally, by tracing the computations and timing of each operation, we generate computational graphs and eliminate incompatible candidates via an estimation process. For a more comprehensive analysis of this method, readers are encouraged to refer to \cite{hong20200wn}.



We deploy a memory fault injection technique to recover the DNN parameters as described in \cite{rakin2021deepsteal}. First, we occupy the majority of memory using the memory exhaustion technique using mmap with MAP\_POPULATE flag. This triggers the OS to move other data (including victim application pages) out of main memory so that the victim pages get evicted from the main memory to the swap space. In the second step, we release pages in order to make the victim pages get relocated into them. In other words, we create a list of potential pages for the victim to occupy as an aggressor during the double-sided rowhammer. Once the victim runs the process again, the victim's pages will be placed in those predetermined locations to create a suitable memory layout for rowohammer leakage.
To achieve this we take advantage of page-frame cache structure called per-cpu pageset. Using this structure we determine the victim page placement with high accuracy based on the releasing order in the second step \cite{yao2020deephammer}. Finally, after the execution of rowhammer, we examine the vulnerable cell changes to find out the value of the bits in the victim row. A similar explanation of the parameter extraction was discussed in \cite{rakin2021deepsteal}. Note that during the parameter recovery process, the values of the victim rows do not change; thus, no accuracy degradation is happening during this step and the attack is performed in stealth. This process is done multiple times until we have recovered enough information about our parameters. Higher recovery rates lead to  more iterations and longer extraction processes. It should be noted that the relationship between time and recovery rate is not linear; higher recovery rates require increasingly longer periods of time. For instance, extracting 90\% of the parameters takes approximately twice as long as extracting 70\% of them.


\section{Experimental Results}\label{sec:results}
In this section, extensive experiments are conducted to evaluate the efficiency of B3FA and its comparison with the state-of-the-art methods. We tested different recovery rates from 90\% to 60\% to evaluate the parameter recovery on different settings such as number of attack rounds and leakage positions of the memory pages. Moreover, the performance of B3FA on several weight representations (quantization bit-widths) and different reconstruction methods are studied. 

\subsection{Results on Different DNNs and Datasets}


We test B3FA on two datasets and three different DNN architectures to evaluate its efficiency of finding vulnerable bits and crushing the model's functionality. We evaluate our approach with different recovery rates to observe the impact of CZR on the attack performance. Since the experiment time grows significantly with the number of bit-flips, we only present the accuracy drop up to 100 bit-flip injections.

Fig. \ref{fig:aveAcc} shows B3FA results on three different DNN architectures. B3FA was able to destroy the functionality of every given model on both datasets with different recovery rates. The higher the recovery rate, the bigger the accuracy drop. Also, B3FA performs better on smaller architectures. This is because, in the same recovery rate, smaller networks have less unrecovered bits and thus, less wrong predicted bits. For instance, a ResNet50 model trained on CIFAR-100 with the recovery rate of 60\% could only decrease the model accuracy by 35\% (from 75\% to around 40\%) after 100 bit-flips (which is still a huge accuracy drop though). For the smaller  MobileNetV2 network trained on CIFAR-100, the accuracy drop is around 62\% (from 70\% to around 8\%) after 100 bit-flips with the same recovery rate (60\%).

 All of the results show a significant decrease in the model accuracy in the first 10 to 20 bit-flips. In some cases, the accuracy even falls below 10\%. For example, for the MobileNetV2 trained on CIFAR-100, B3FA is able to decrease the accuracy of the model from 69.84\% to 8.1\% with only 20 bit-flips with a recovery rate of 80\%. Since CZR does not predict the unrecovered value with a 100\% accuracy, in the partially extracted setting, the FL2R can produce different results and leads to variance. Note that the variance of model accuracy in Fig. \ref{fig:aveAcc} is directly related to the recovery rates; Higher recovery rates relate to lower variance and lower recovery rates relate to higher variance.

\subsection{Comparison to Traditional White-box Attacks}\label{sec:white-box}
We compare our proposed FL2R approach with the previous state-of-the-art white-box method \cite{rakin2019bit} in terms of the vulnerable bit-ranking efficiency. Since the work in \cite{rakin2019bit} assumes white-box settings, it only works while having a fully recovered DNN. Thus, to use it in a semi-black-box setting, we have to use a method for the prediction of unextracted bits; We use a random bit generation for reconstruction of unrecovered parameter bits. Furthermore, we assume \cite{rakin2019bit} has access to a portion of the dataset whereas FL2R ranks the vulnerable bits blindly without having access to the dataset. 

\begin{figure}[!tb]
	\begin{center}
	    \vspace{-0.2in}
		\includegraphics[width = 0.4\textwidth]{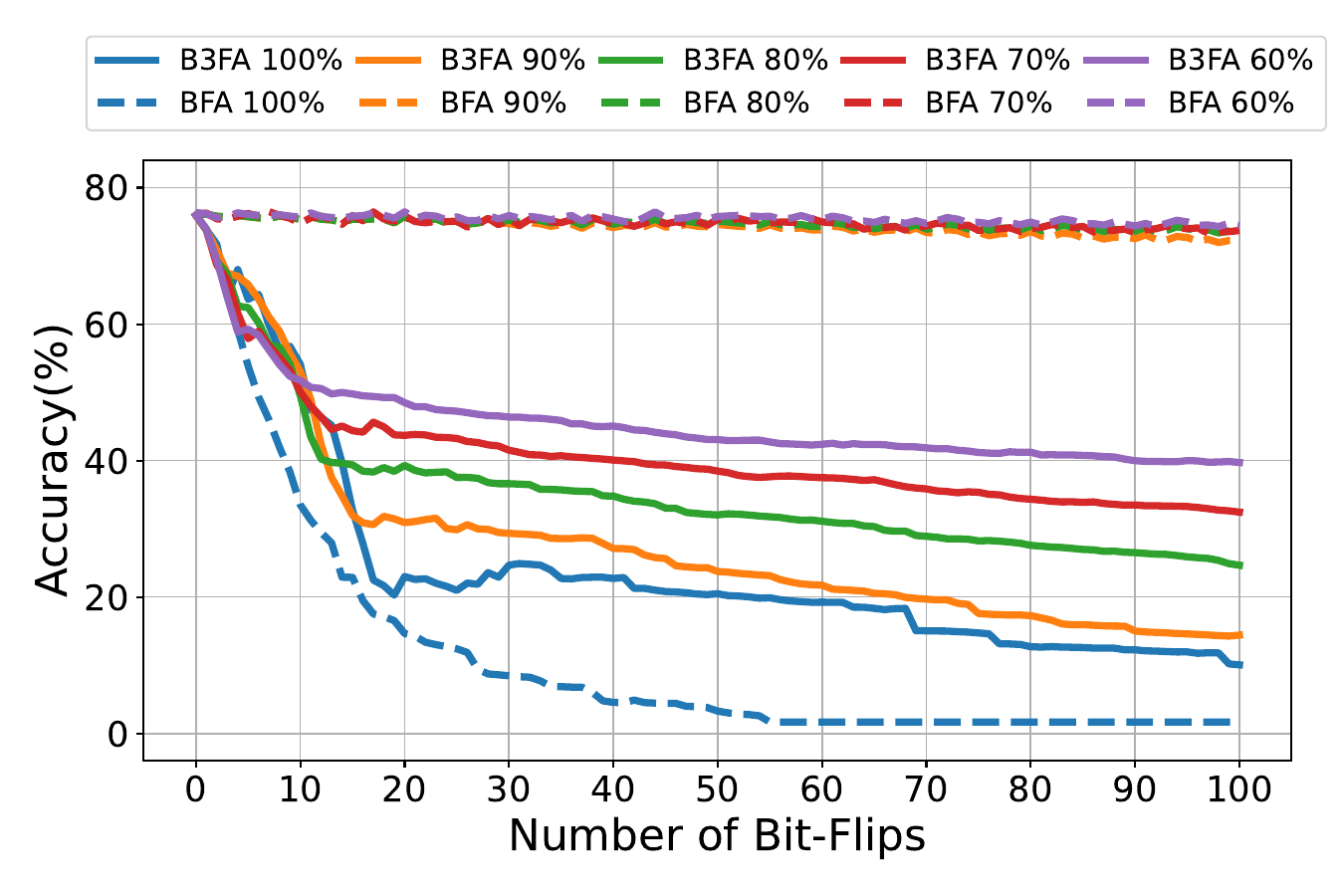}
		\vspace{-0.1in}
		\caption{Comparison of B3FA and BFA \cite{rakin2019bit} on ResNet50 with CIFAR-100 at various recovery rates.}
		\vspace{-0.1in}
		\label{fig:bfa-vs-b3fa}
	\end{center}	
\vspace{-0.1in}
\end{figure}

Both architecture and weight recovery processes may introduce imperfections and errors, which are considered when evaluating the effectiveness of the proposed BFA. An important aspect of our approach is its ability to operate effectively even in scenarios where errors are present in the recovered model. Fig. \ref{fig:bfa-vs-b3fa} illustrates the results of attacking a ResNet50 model trained on CIFAR-100 using BFA \cite{rakin2019bit} and our B3FA. Despite having access to the dataset, BFA fails to mount successful attacks on partially recovered models in all settings. Even at a recovery rate of 90\%, BFA can only cause a maximum accuracy drop of 2\% after 100 bit-flips. However, with only 20 bit-flips, B3FA induces a 44\% accuracy drop at the same recovery rate. Remarkably, even in the 100\% recovery rate (full parameter recovery), while BFA has access to the dataset and our B3FA does not, our B3FA results are still comparable to BFA. Thus, B3FA emerges as a strong candidate for attacks in white-box settings.

\begin{figure}[!tb]
	\begin{center}
		\vspace{-0.1in}
		\includegraphics[width = 0.4\textwidth]{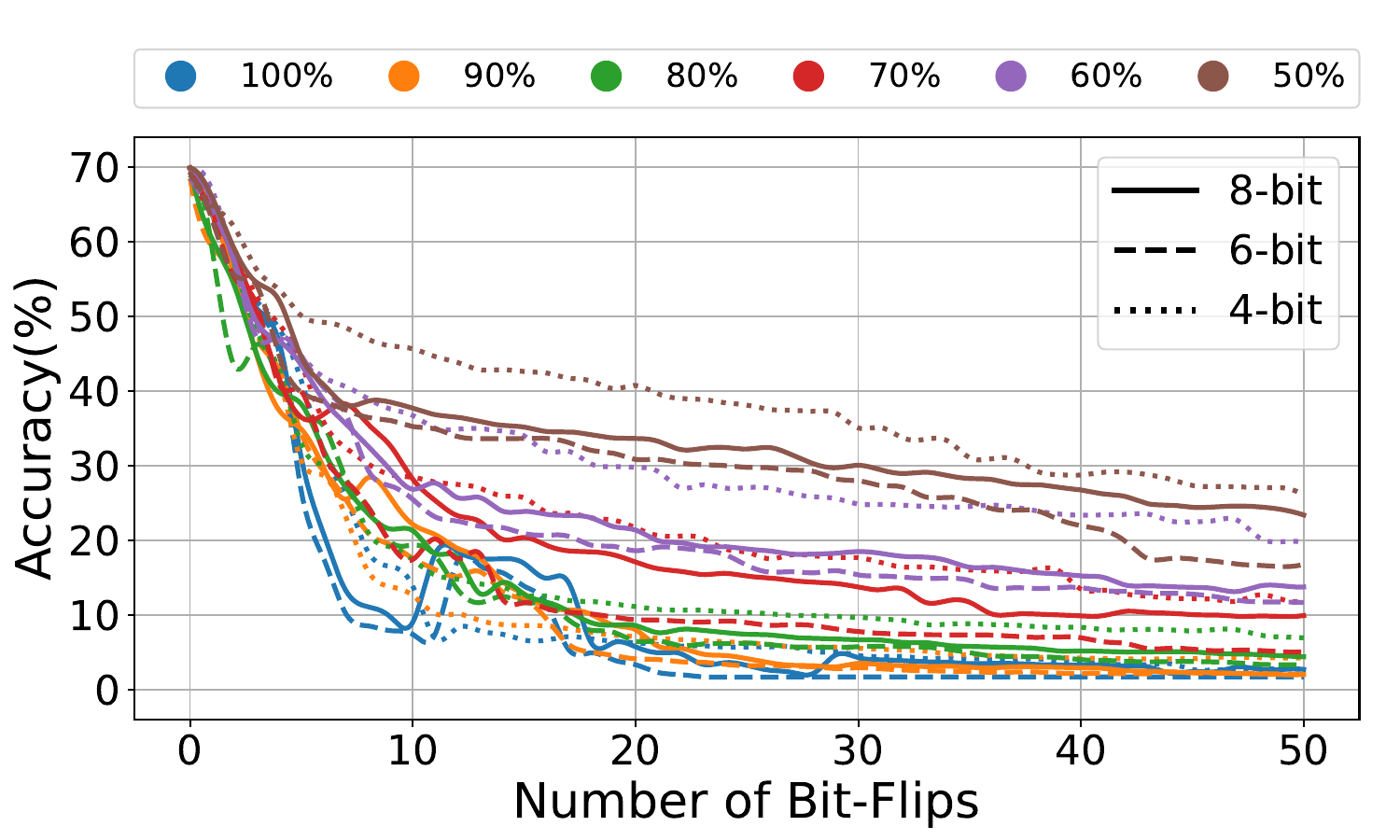}
		\vspace{-0.1in}
		\caption{ B3FA results on MobileNetV2 with CIFAR-100 for different quantization bit-widths (8, 6, 4). Each configuration tested 10 times per recovery rate; averages shown.}
		\label{fig:qb3fa}
		 \vspace{-0.1in}
	\end{center}	
\end{figure}
\begin{figure}[!tb]
	\begin{center}
	    \vspace{-0.2in}
		\includegraphics[width = 0.38\textwidth]{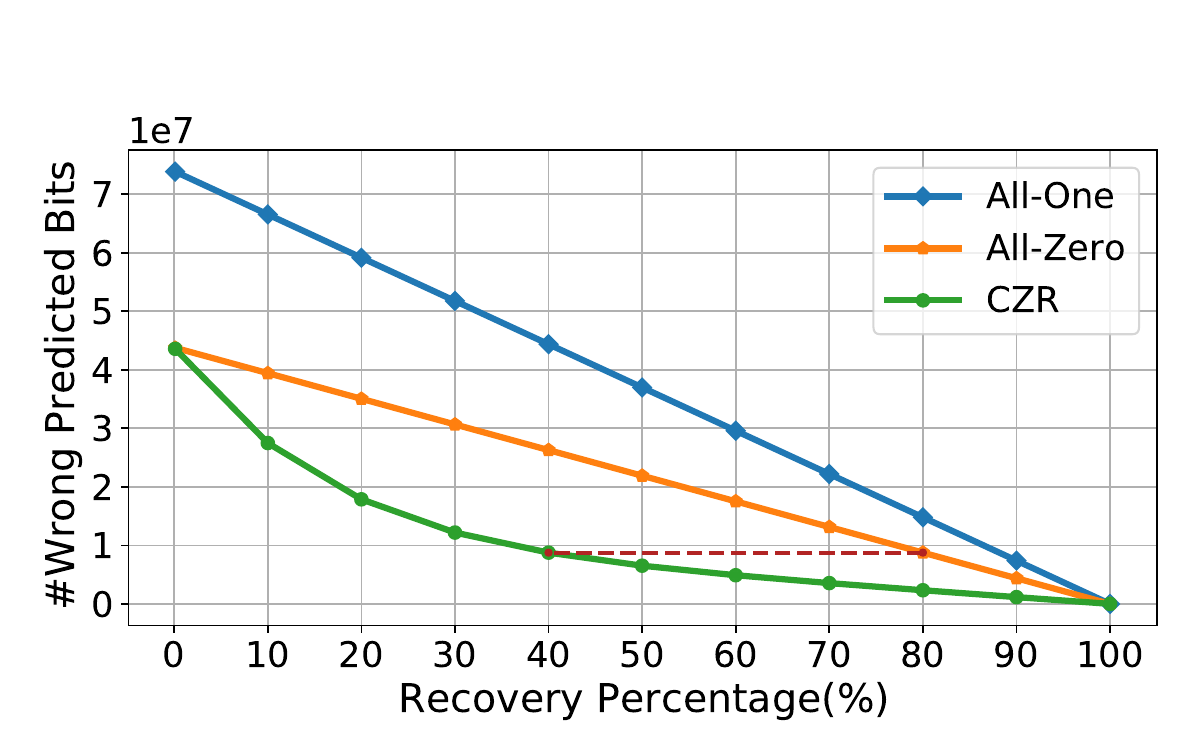}
		\vspace{-0.1in}
		\caption{Comparison of different reconstruction methods with different Recovery Percentages evaluated on a VGG16 model trained on CIFAR10.}
		\label{fig:recons}
	\end{center}	
\vspace{-0.1in}
\end{figure}

\subsection{Results on Various Weight Representation}
DNNs are typically stored in memory using either floating-point (float) or fixed-point (quantized) formats. However, the floating-point format is not very resilient to fault; even a single bit-flip in the Most Significant Bit (MSB) of a parameter can lead to significant changes in the output. This means that a random bit-flip attack could occur, leading to unpredictable results \cite{rakin2019bit}. As a result, this section focuses on the use of quantized DNNs, which not only provide a more compressed representation of the model, but also result in a high level of bit-flip resistance.
It is woth to mention that The existing literature has shown that the BFA attack on floating-point models lacks challenge and does not require a vulnerability search (i.e. simply randomly flipping the exponent part of floating-point weights can easily disrupt DNN functionality.).


To show the performance of B3FA on different quantization bit-widths, we evaluate our approach on 8-bit, 6-bit, and 4-bit quantized MobileNetV2 trained on CIFAR-100. 
Since the main target of B3FA is the sign bit of the parameters, the change in parameter values will be almost equal in different bit-widths. 
Thus, As illustrated in Fig. \ref{fig:qb3fa}, we can see that B3FA is able to successfully attack all networks with different bit widths and decrease 
their accuracy to below 10\% for the recovery rates of 100\%, 90\%, and 80\%. For the recovery rates of 70\%, 60\%, and 50\%, B3FA is able to to cause a huge accuracy drop as well, and decrease the accuracy of DNN to below 30\%. Although the attack performance was nearly similar in high recovery rates. In lower recovery rates, 6-bit quantization consistently yielded the highest performance compared to 8 and 4-bit ones.
\subsection{Study of Different Reconstruction Methods}
In order to evaluate the efficiency of the proposed CZR, various reconstruction algorithms are explored to ensure that partially extracted parameters from memory can be accurately reconstructed to closely resemble the original parameter values. These algorithms include the All-Zeros method (in which unrecovered bits are assigned a value of zero), the All-Ones method (in which unrecovered bits are assigned a vlue of one). As shown in Fig \ref{fig:recons}, CZR is much more accurate than the other methods and has less number of wrong predicted bits in each recovery rate. For instance, the accuracy of CZR method with 40\% recovery rate is same as the accuracy for All-Zero with 90\% recovery rate. This is because the All-Zeros method has a high accuracy for reconstruction of positive values, and the All-Ones method has a high accuracy for reconstruction of negative values. However, since CZR is a combination of the other two methods, it benefits from the high performance of All-Zeros for the positive values, and All-Ones for the negative values.

\section{Conclusion and Discussion}
\label{sec:disc&con}
In this section, we discuss defense strategies against B3FA and provide a conclusion.

\subsection{Defense Strategies against B3FA}
There are various defense mechanisms that can be used to protect DNNs from adversarial parameter attacks. Every protection mechanism that could interrupt the chain of steps of the attack will be efficient. It is worth noting that these defence mechanisms require additional runtime and incur high computational overhead and  the cost of the mechanism should be considered.

One category of these mechanisms involves implementing additional selective safety protocols. For example, the Hashtag approach \cite{javaheripi2021hashtag} proposed a sensitivity measure to identify the most vulnerable DNN layers and encode the parameters in those layers using a low-collision hash function in order to validate the integrity of the model. This sensitivity measure is gradient-based, meaning that it identifies the layers with the most vulnerable parameters that a gradient-based bit-flip attack aims to perturb. However, in B3FA which is a magnitude-based approach, a different set of vulnerable bits and vulnerable layers are exposed to bit-flip attack. This is mainly because the magnitude and the gradient of a parameter does not have any direct relation (e.g, a high value parameter can have a low gradient and vice versa). Therefore, applying the concept of Hashtag and developing a new magnitude-based sensitivity measure can be an effective defense strategy against B3FA.

Another category of defense strategies involves changing the model either through binarization or fine-tuning \cite{he2020defending}. These approaches can be applied to mitigate the effect of B3FA by lowering the scaled $\mathcal{L}_2$-norm of vulnerable filters and equalizing all the DNN filters in order to complicate the vulnerable bit-selection process. 


\subsection{Conclusion}
As DNNs continue to achieve great success in various areas, they are increasingly being deployed in critical systems such as personal identity recognition systems and autonomous vehicles. However, this also creates significant security concerns regarding their deployment. While researchers have already investigated various types of bit-flip attacks on DNNs and proposed countermeasures, this paper introduces a new type of semi-black-box bit-flip attack on DNNs. We propose B3FA, a practical attack that can perform adversarial attacks on DNN parameters under the assumptions of having no data and partially extracted DNN parameters. Our attack highlights the vulnerability of DNN parameters to bit-flips in real-life scenarios, which is a major concern for the security and reliability of critical systems that rely on these models. The results of our study underscore the need for developing new and more robust defense strategies to prevent such attacks and safeguard against their potentially disastrous consequences.
The attack is based on recent advances in model parameter and architecture extractions. Our attack, B3FA, is able to demolish the functionality of every examined DNN and dataset, highlighting the vulnerability of DNN parameters to bit-flips in real-life scenarios using practical attacks. The results of our study demonstrate the urgent need for new defense strategies to address this vulnerability and prevent future attacks.

\section{Acknowledgements}
The authors would like to thank Dr. Raji and Dr. Fang for their valuable insights and review comments. We acknowledge the support from the Natural Sciences and Engineering Research Council (NSERC) of Canada, This work is funded by the Natural Sciences and Engineering Research Council of Canada NSERC https://www.nserccrsng.gc.ca/ under Grant No. NETGP485577-15 NSERC (COHESA project) and 341516 NSERC (RGPIN), along with in-kind support from AMD and Intel/Altera.

\bibliographystyle{IEEEtran}
\bibliography{DeepSteal}

\begin{thebibliography}{10}
\providecommand{\url}[1]{#1}
\csname url@samestyle\endcsname
\providecommand{\newblock}{\relax}
\providecommand{\bibinfo}[2]{#2}
\providecommand{\BIBentrySTDinterwordspacing}{\spaceskip=0pt\relax}
\providecommand{\BIBentryALTinterwordstretchfactor}{4}
\providecommand{\BIBentryALTinterwordspacing}{\spaceskip=\fontdimen2\font plus
\BIBentryALTinterwordstretchfactor\fontdimen3\font minus \fontdimen4\font\relax}
\providecommand{\BIBforeignlanguage}[2]{{%
\expandafter\ifx\csname l@#1\endcsname\relax
\typeout{** WARNING: IEEEtran.bst: No hyphenation pattern has been}%
\typeout{** loaded for the language `#1'. Using the pattern for}%
\typeout{** the default language instead.}%
\else
\language=\csname l@#1\endcsname
\fi
#2}}
\providecommand{\BIBdecl}{\relax}
\BIBdecl

\bibitem{kim2014flipping}
{Kim et al.}, ``Flipping bits in memory without accessing them: An experimental study of dram disturbance errors,'' \emph{SIGARCH 2016}, vol.~42, no.~3, pp. 361--372.

\bibitem{yao2020deephammer}
{Yao et al.}, ``$\{$DeepHammer$\}$: Depleting the intelligence of deep neural networks through targeted chain of bit flips,'' in \emph{USENIX Security 2020}, pp. 1463--1480.

\bibitem{rakin2019bit}
{Rakin et al.}, ``Bit-flip attack: Crushing neural network with progressive bit search,'' in \emph{ICCV 2019}, pp. 1211--1220.

\bibitem{rakin2021deepsteal}
------, ``Deepsteal: Advanced model extractions leveraging efficient weight stealing in memories,'' in \emph{SP 2022}.\hskip 1em plus 0.5em minus 0.4em\relax IEEE, pp. 1157--1174.

\bibitem{agoyan2010flip}
{Agoyan et al.}, ``How to flip a bit?'' in \emph{IOLTS 2010}, pp. 235--239.

\bibitem{liu2017fault}
{Liu et al.}, ``Fault injection attack on deep neural network,'' in \emph{ICCAD 2017}, pp. 131--138.

\bibitem{hong2019terminal}
{Hong et al.}, ``Terminal brain damage: Exposing the graceless degradation in deep neural networks under hardware fault attacks,'' in \emph{USENIX Security 2019}, pp. 497--514.

\bibitem{zhao2019fault}
P.~Zhao, S.~Wang, C.~Gongye, Y.~Wang, Y.~Fei, and X.~Lin, ``Fault sneaking attack: A stealthy framework for misleading deep neural networks,'' in \emph{Proceedings of the 56th Annual Design Automation Conference 2019}, 2019, pp. 1--6.

\bibitem{ghavami2022stealthy}
{Ghavami et al.}, ``Stealthy attack on algorithmic-protected dnns via smart bit flipping,'' in \emph{ISQED 2022}.\hskip 1em plus 0.5em minus 0.4em\relax IEEE, 2022, pp. 1--7.

\bibitem{rakin2021t}
R.~et~al., ``T-bfa: Targeted bit-flip adversarial weight attack,'' \emph{IEEE Transactions on Pattern Analysis and Machine Intelligence}, 2021.

\bibitem{bai2021targeted}
J.~Bai, B.~Wu, Y.~Zhang, Y.~Li, Z.~Li, and S.-T. Xia, ``Targeted attack against deep neural networks via flipping limited weight bits,'' \emph{arXiv preprint arXiv:2102.10496}, 2021.

\bibitem{ghavami2021bdfa}
{Ghavami et al.}, ``Bdfa: A blind data adversarial bit-flip attack on deep neural networks,'' in \emph{DSD 2022.}\hskip 1em plus 0.5em minus 0.4em\relax IEEE.

\bibitem{ribeiro2015mlaas}
{Ribeiro et al.}, ``Mlaas: Machine learning as a service,'' in \emph{ICMLA 2015}, pp. 896--902.

\bibitem{hong20200wn}
{Sanghyun Hong et al.}, ``How to 0wn nas in your spare time,'' \emph{ICLR 2020}.

\bibitem{hu2020deepsniffer}
{Hu et al.}, ``Deepsniffer: A dnn model extraction framework based on learning architectural hints,'' in \emph{ASPLOS 2020}, pp. 385--399.

\bibitem{kwong2020rambleed}
{Kwong et al.}, ``Rambleed: Reading bits in memory without accessing them,'' in \emph{2020 IEEE Symposium on Security and Privacy (SP)}.\hskip 1em plus 0.5em minus 0.4em\relax IEEE, 2020, pp. 695--711.

\bibitem{razavi2016flip}
{Razavi et al.}, ``Flip feng shui: Hammering a needle in the software stack,'' in \emph{USENIX Security 2016}, pp. 1--18.

\bibitem{seaborn2015exploiting}
M.~Seaborn and T.~Dullien, ``Exploiting the dram rowhammer bug to gain kernel privileges,'' \emph{Black Hat}, vol.~15, p.~71, 2015.

\bibitem{yan2020cache}
{Yan et al.}, ``Cache telepathy: Leveraging shared resource attacks to learn dnn architectures,'' in \emph{USENIX Security 2020}, pp. 2003--2020.

\bibitem{batina2019csi}
{Batina et al.}, ``$\{$CSI$\}$$\{$NN$\}$: Reverse engineering of neural network architectures through electromagnetic side channel,'' in \emph{USENIX Security 2019}, pp. 515--532.

\bibitem{breier2021sniff}
{Breier et al.}, ``Sniff: reverse engineering of neural networks with fault attacks,'' \emph{IEEE Transactions on Reliability}, 2021.

\bibitem{jagielski2020high}
{Jagielski et al.}, ``High accuracy and high fidelity extraction of neural networks,'' in \emph{USENIX security 2020}, pp. 1345--1362.

\bibitem{he2018soft}
{He et al.}, ``Soft filter pruning for accelerating deep convolutional neural networks,'' \emph{arXiv preprint arXiv:1808.06866}, 2018.

\bibitem{gholami2021survey}
{Gholami et al.}, ``A survey of quantization methods for efficient neural network inference,'' \emph{arXiv preprint arXiv:2103.13630}, 2021.

\bibitem{lin2016fixed}
{Lin et al.}, ``Fixed point quantization of deep convolutional networks,'' in \emph{ICML 2016}, pp. 2849--2858.

\bibitem{sandler2018mobilenetv2}
{Sandler et al.}, ``Mobilenetv2: Inverted residuals and linear bottlenecks,'' in \emph{CVPR 2018}, pp. 4510--4520.

\bibitem{simonyan2014very}
{Simonyan et al.}, ``Very deep convolutional networks for large-scale image recognition,'' \emph{arXiv preprint arXiv:1409.1556}, 2014.

\bibitem{he2016deep}
{He et al.}, ``Deep residual learning for image recognition,'' in \emph{CVPR 2016}, pp. 770--778.

\bibitem{CIFAR-10}
{Alex Krizhevsky et al.}, ``Cifar-10 (canadian institute for advanced research),'' \emph{https://www.cs.toronto.edu/~kriz/cifar.html}, 2009.

\bibitem{paszke2019pytorch}
{Paszke et al.}, ``Pytorch: An imperative style, high-performance deep learning library,'' in \emph{NeurIPS 2019}.

\bibitem{javaheripi2021hashtag}
{Javaheripi et al.}, ``Hashtag: Hash signatures for online detection of fault-injection attacks on deep neural networks,'' in \emph{ICCAD 2021}, pp. 1--9.

\bibitem{he2020defending}
Z.~He, A.~S. Rakin, J.~Li, C.~Chakrabarti, and D.~Fan, ``Defending and harnessing the bit-flip based adversarial weight attack,'' in \emph{Proceedings of the IEEE/CVF Conference on Computer Vision and Pattern Recognition}, 2020, pp. 14\,095--14\,103.

\end{thebibliography}



\end{document}